\documentstyle[11pt,IAU207_pasp,twoside,psfig]{article}

\def\edcomment#1{\iffalse\marginpar{\raggedright\sl#1\/}\else\relax\fi}
\reversemarginpar

\begin{document}
\title{Spectroscopy of Globular Clusters in NGC 1399 - A Progress Report}
 \author{T. Richtler, B. Dirsch, D. Geisler}
\affil{Grupo de Astronom\'{\i}a, Departamento de F\'{\i}sica, Universidad de Concepci\'on, Casilla 160-C, Concepci\'on, Chile}
\author{K. Gebhardt}
\affil{Department of Astronomy, University of Texas at Austin, USA}
\author{M. Hilker, L. Infante, D. Minniti, M. Rejkuba}
\affil{Departamento de Astronom\'{\i}a y Astrof\'{\i}sica, P. Universidad Cat\'olica, Vicu\~na Mackenna 4860, Santiago 22, Chile}

\author{J. C. Forte}
\affil{Facultad de Ciencias Astronomicas y Geof\'{\i}sicas, Universidad Nacional de La Plata, 1900 La Plata, Argentina}

\author{S. Larsen}
\affil{Lick Observatory, University of California at Santa Cruz, USA}
\author{E. K. Grebel}
\affil{MPI f\"ur Astronomie, K\"onigstuhl 17, D-69117, Heidelberg, Germany}
\author{V. Alonso}
\affil{Observatorio astronomico Cordoba, Laprida 854, 5000 Cordoba, Argentina}
\begin{abstract}
We performed spectroscopy of globular clusters associated with NGC 1399 and measured radial velocities of more than 400 clusters, the largest sample ever obtained for dynamical studies. In this
progress report, we present the sample and the first preliminary results. Red and blue clusters
have slightly different velocity dispersions in accordance with their different density profiles.
Their velocity dispersions remain constant with radial distance, which differs from earlier work.  
\end{abstract}

\section{Introduction}

We performed spectroscopy of globular clusters in NGC 1399, the central cD-galaxy in the Fornax cluster, with a double intention: firstly, we want to use globular clusters as probes of the gravitational potential of this galaxy and
 secondly, we
want to investigate whether there are kinematical differences among subpopulations of clusters with different metallicities. Both X-ray analyses (Jones et al.
1997, Ikebe et al. 1996) and previous studies of globular cluster velocities (Grillmair et al. 1994, Kissler-Patig et al. 1999) indicated that NGC 1399 should possess a massive dark matter halo. However, the sizes of earlier
samples of cluster velocities were too small to allow a definitive dynamical analysis. We therefore aimed at
obtaining enough radial velocities to determine both 
the potential of the galaxy and the phase space distribution of clusters. In this contribution, we report on our progress in this project and present the first preliminary results.

\section{Observations and Data Analysis }

We pre-selected cluster
candidates by photometry in Washington C and Cousins R of MOSAIC data from the 4m telescope at CTIO (see Dirsch et al. 2001, these proceedings). Our candidates
cover a magnitude range of 
  20 $<$ R $<$ 23 and a color range of 0.7 $<$ C-R $<$ 2.2. An additional
criterion was the ''stellarity''-index returned by SExtractor to discriminate against background galaxies. As the radial velocities later on showed, the
 success rate of selecting clusters and not stars or galaxies,
was very high, about 95 \%.
The spectroscopic observations have been performed in the period 30.11-2.12.2000 at the VLT
(Kuyen) with FORS2 and the Mask Exchange Unit (MXU). The grism was 600B, giving
a spectral resolution of about 2.5 {\AA}, based on the widths of the arc lines.
 The spectral range which could be used was 3800 to 6000 {\AA}, depending on the slit
position in the mask and the signal-to-noise. To be flexible with respect to the object selection,
 we decided to set sky slits independently from the objects slits. For most of the slits, the size was $1\times 2 \arcsec$. The number of slits on a mask varied  between 100 and 120. We exposed 13 masks (exposure times were either 45 min or 2$\times$45 min)
and
obtained spectra of about 500 cluster candidates and many miscellaneous objects like
point sources outside our color and magnitude limits and galaxies, as well as
hundreds of sky spectra at varying galactocentric radii. The range of
radial distances from the galaxy center was 2$\arcmin$ $<$ R $<$ 8$\arcmin$, corresponding to 11 kpc $<$ R $<$ 44 kpc with an adopted distance of 19 Mpc. The extraction of the
spectra was done with "apextract" under IRAF. The wavelength
calibration was very accurate, so that subtracting an independently calibrated
sky from an object worked well. Small zeropoint differences of the order 0.5 {\AA} between the masks were adjusted by matching the sky lines.

We measured radial velocities by cross-correlating a suitable part of the spectrum with a template spectrum, which we chose to be that of NGC 1396. In general
the interval 4500 - 5500 {\AA} gave the highest correlation peaks. The errors,
 which were returned by the correlation software (``fxcor'' under IRAF), varied between 10 km/s for our brightest objects and more than 100 km/s for the
 faintest ones (when a reasonable correlation peak could be found). We verified that these errors are indeed resonable by comparing those objects which were
found in two different masks (about 25).   

We excluded objects with velocity errors larger than 100 km/s. Moreover,
we excluded 
 a few cluster candidates with radial velocities smaller than 500 km/s and larger than 2500 km/s.
After that our sample now consists of 369 clusters. The final sample will have more than 400.
 We expect
 that  many miscellaneous stellar-like objects will still turn out to be globular clusters, but we already now have  the largest sample ever obtained for a dynamical analysis of a globular cluster system.

\section{Preliminary Results}

Fig. 1 shows in the upper panel the entire sample of 369 velocities vs. the angular galactocentric distance in arcmin.   

\begin{figure}
\centerline{\vbox{
\psfig{figure=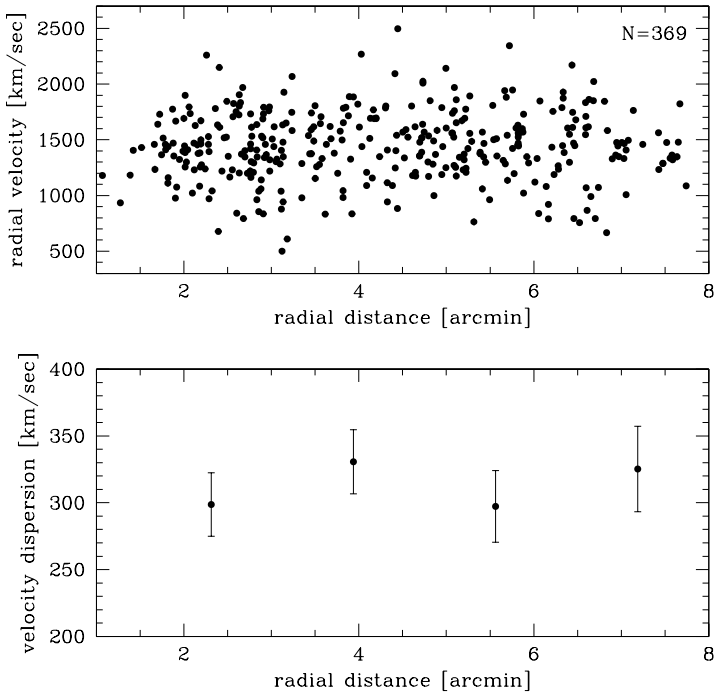,width=10cm}
}}
\caption{The upper panel shows the error-selected ($<$ 100 km/s) and velocity-selected (500 km/s $< v_r <$ 2500 km/s) sample of globular cluster velocities.
In the lower panel we plot the measured velocity dispersions in four radial bins.
We do not see an increase of the velocity dispersion with radius, which
contrasts with earlier work. 
}
\end{figure}

 The measured (projected)
velocity dispersion of the entire sample is $310 \pm 20$ km/s. The lower panel gives the
dispersions in four radial bins.  It remains
constant within the errors over the whole radial range (at this moment, the dispersion is simply calculated as the dispersion of a Gaussian and has not been 
estimated with more 
sophisticated statistical tools). This is in disagreement with
earlier work (Kissler-Patig et al. 1999) which suggested a considerable increase  between
2$\arcmin$ and 6$\arcmin$.

In the bimodal color distribution of clusters (see Dirsch et al. 2001, these proceedings), we use a color of $\rm C - R$ = 1.4 to separate  the metal-poor from the metal-rich clusters. 
 The metal-poor (190 objects) and the metal-rich (179 objects) subsamples show slightly 
different velocity dispersions of $330 \pm 18$ km/s and $291 \pm 16$ km/s, respectively, and again do
 not indicate a change with radial distance. These values decrease
 insignificantly to $326\pm18$ and $286\pm16$, if we subtract the contribution of our
 mean error of 50 km/s to the velocity dispersion.

 Fig.2 shows the mean dispersions  for the blue and the red clusters and that
their difference remains stable if we select objects with progressively
smaller velocity errors. 
\begin{figure}[t]
\centerline{\vbox{
\psfig{figure= 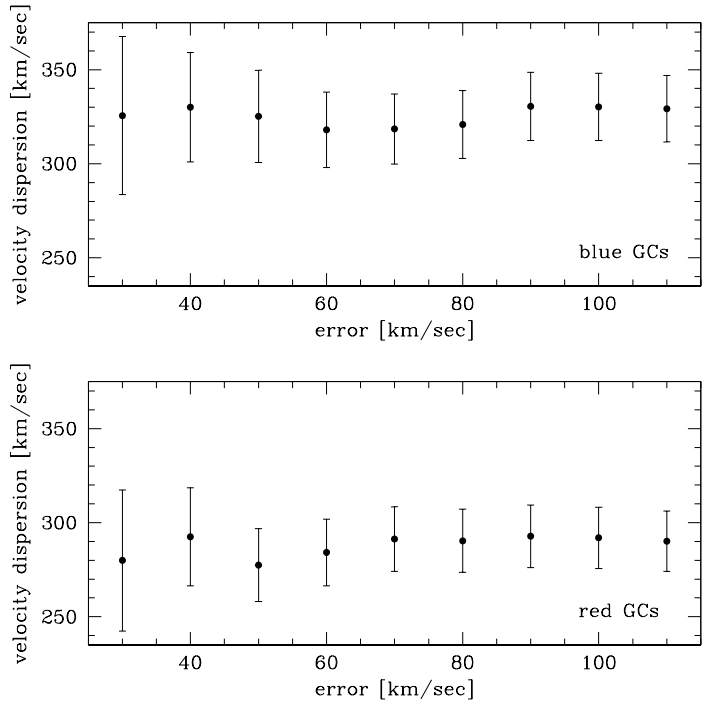,width=10cm}
}}
\caption{
 The ordinate
gives the velocity dispersion of the error selected samples of red and blues clusters,
respectively. The abcissa is the maximum error which was allowed to calculate
the respective dispersion.  This plot demonstrates that the difference between the velocity dispersions of the red and the blue clusters stays if we select clusters with
progressively smaller velocity errors.}
\end{figure}

We do not see any
 significant rotation for most of the cluster population except perhaps for the outer
metal-poor clusters, for which a marginal rotation signature might be present.

The dynamical analysis, in which we intend to apply axisymmetric, three-integral models (Gebhardt et al. 2000) will supersede the following exercise. As a first approach, we assume spherical symmetry (NGC 1399 is slightly elliptical) and neglect rotation. Then the radial Jeans-equation  reads
$$ \frac{G \cdot M(r)}{r} = - \sigma_{r}^2 \cdot (\frac{d\ln\rho}{d\ln r} 
+ \frac{d\ln\sigma_{r}}{d\ln r} + 2 \beta) $$
where G is the constant of gravitation, r the galactocentric distance, M(r) the
mass contained within r, $\sigma_r$ the radial component of the velocity dispersion, $\rho$(r)
the density profile of clusters, $\beta = 1- \frac{\sigma_{\Theta}^2}{\sigma_{r}^2}$ with 
$\sigma_{\Theta}$ being the tangential velocity dispersion.
Unless $\beta$ is large, a radially constant {\it projected} velocity dispersion implies a constant
$\sigma_r$ and $\frac{d~ln\sigma_{r}}{d~ln~r}= 0$.   
Within our radial range, the red and the blue clusters show different density profiles (Dirsch et al. 2001, these proceedings), $\frac{d~ln\rho}{d~ln~r} \sim$ -2.5$\pm$0.1 and -1.8$\pm$0.1,
 respectively. They must trace the same mass, so it is interesting that the difference of the density profiles could account completely
for the difference in the velocity dispersions, if $\beta$ would be zero.

A handy formula for the mass inside a radius r is ($\beta = 0$) 
$$ M[M_{\odot}] = 2 \cdot 10^{10} \cdot r[kpc] \cdot (\frac{\sigma_r^2}{300 km/s})^2 \cdot \alpha$$
where $\alpha$ is the slope of $\rho$(r) and where a distance of 19 Mpc has been assumed.
The mass inside 10 kpc is $4.2 \cdot 10^{11} M_{\odot}$ and thus inside 40 kpc $1.7 \cdot 10^{12} M_{\odot}$, which is in good agreement with the values given by Jones et al. (1997) and somewhat higher
than the values found in Ikebe et al. (1996), who moreover find a shallower dependence of M(r) than
$\rm M(r) \sim  r$.  

A radially constant $\sigma_r$ and thus a constant circular velocity is an interesting analogy
 to disk galaxies. NGC 1399 is an example of a
spheroidal galaxy, where a dark halo and luminous component might ''conspire''
 to show a constant circular velocity.  Since this is the first time that the
dynamical analysis  can be performed out to 4 effective radii, one can expect
to further constrain the properties of the dark halo.

\section{Reference List}

\begin{quote}
Dirsch B, Geisler D., Richtler T., Forte J.C. 2001, these proceedings\\
Gebhardt K. et al. 2000, AJ, 119, 1157\\
Grillmair, C. et al. 1994, ApJ, 422, L9\\
Ikebe, Y. et al. 1996, Nature, 379, 427 \\
Jones, C. et al. 1997,  ApJ 482, 143 \\
Kissler-Patig M. et al. 1999, AJ, 117, 1206\\

\end{quote}

\end{document}